\documentclass[12pt]{article}

\usepackage{amsmath}
\usepackage{amsthm}

\def\fnote#1#2{\begingroup\def\thefootnote{#1}\footnote{#2}
    \addtocounter{footnote}{-1}\endgroup}

\def\Email{makoto.natsuume@kek.jp}
\def\asano{asano@hep-th.phys.s.u-tokyo.ac.jp}

\newcommand{\bra}[1]{\mbox{$\langle #1 |$}}
\newcommand{\ket}[1]{\mbox{$| #1 \rangle$}}
\newcommand{\norm}[2]{\mbox{$\langle #1 | #2 \rangle$}}

\newcommand{\sig}[1]{\mbox{sign}(#1)}
\newcommand{\dims}[1]{\mbox{dim}(#1)}
\newcommand{\tr}{\mbox{tr}\,}

\newcommand{\defi}{\stackrel{\rm def}{=}}

\newcommand{\eq}[1]{(\ref{eq:#1})}
\newcommand{\cond}[1]{\quad\mbox{#1}}

\newcommand{\hN}{\hat{N}^g}
\newcommand{\hQ}{\hat{Q}}
\newcommand{\ta}{\tilde{a}}
\newcommand{\tb}{\tilde{b}}

\newtheorem{lemma}{Lemma}[section]
\newtheorem{theorem}{Theorem}[section]

\def\IR{\relax{\rm I\kern-.18em R}}

\begin{document}

\pagestyle{empty}

\begin{flushright}
        KEK-TH-688, UT-886\\
        hep-th/0005002\\
\end{flushright}

\vspace{18pt}

\begin{center}
{\large \bf The no-ghost theorem for string theory} \\ \vspace{4pt}
{\large \bf in curved backgrounds with a flat timelike direction}

\vspace{16pt}
Masako Asano $^{1}$ \fnote{\dag}{\asano} and Makoto Natsuume $^{2}$
\fnote{*}{\Email}

\vspace{16pt}

{\sl $^{1}$ Department of Physics\\
University of Tokyo\\
Hongo, Bunkyo-ku, Tokyo, 113-0033 Japan}

\vspace{8pt}

{\sl $^{2}$ Theory Division\\
Institute of Particle and Nuclear Studies\\
KEK, High Energy Accelerator Research Organization\\
Tsukuba, Ibaraki, 305-0801 Japan}

\vspace{12pt}
{\bf ABSTRACT}

\end{center}

\begin{minipage}{4.8in}
It is well-known that the standard no-ghost theorem can be extended 
straightforwardly to the general $c=26$ CFT on $ \IR^{d-1,1} \times K $, where $2 \leq d \leq 26$ and $ K $ is a compact unitary CFT of appropriate central charge. We prove the no-ghost theorem for $d=1$, {\it i.e.}, when only the timelike direction is flat. This is done using the technique of Frenkel, Garland and Zuckerman.
\end{minipage}

\begin{flushright}
       PACS codes: 11.25.-w, 11.25.Hf, 11.25.Mj\\
       Keywords: no-ghost theorem, BRST quantization,\\
       cohomology, vanishing theorem\\
\end{flushright}

\vfill
\pagebreak

\pagestyle{plain}               
\setcounter{page}{1}    

\baselineskip=16pt

\section{Introduction}

As is well-known, string theory generally contains negative norm states
(ghosts) from timelike oscillators. However, they do not appear as physical
states. This is the famous {\it no-ghost theorem}
\cite{ocq1}-\cite{GM}.\footnote{For the NSR string, see
Refs.~\cite{nsr1}-\cite{FK}.}

When the background spacetime is curved, things are not clear though.
First of all, nonstationary situations are complicated even in field
theory. There is no well-defined ground state with respect to a time
translation and the ``particle interpretation" becomes ambiguous. Moreover,
even for static or stationary backgrounds, it is not currently possible to
show the no-ghost theorem from the knowledge of the algebra alone. In fact,
Refs.~\cite{sl2r} and Ref.~\cite{NS} found ghosts for strings on
three-dimensional anti-de~Sitter space ($AdS_3$)
and three-dimensional black holes, 
respectively. There are references which have proposed resolutions to the
ghost problem  for $AdS_3$ \cite{sl2r2,Evans:1998qu}.
However, the issue is not settled yet and proof for the other backgrounds is still lacking.

Viewing the situation, we would like to ask the converse question: how
general can the no-ghost theorem definitely apply? This is the theme of this
paper.

In order to answer to the question, let us look at the known proofs more
carefully. There are two approaches. The first approach uses the ``old
covariant quantization" (OCQ). Some proofs in this approach use the ``DDF
operators" to explicitly construct the observable Hilbert space to show the
theorem \cite{ocq1,nsr1}. This proof assumes flat spacetime and cannot be
extended to the more general cases. Goddard and Thorn's proof
\cite{ocq2,thorn3} is similar to this traditional one, but is formulated
without explicit reference to the DDF operators. Since it only requires the existence of a flat light-cone vector, the proof can be extended easily to $2 \leq d \leq 26$.

There is another approach using the BRST quantization. These work more
generally. Many proofs can be easily extended to the general $c=26$
CFT on $ \IR^{d-1,1} \times K $, where $2 \leq d \leq 26$ and $K$ corresponds to a compact unitary CFT of central charge $ c_{K} = 26-d $
\cite{KO,FGZ,spiegelglas,GM,Big}. 

On the other hand, the theorem has not been studied in detail when $0<d<2$. The
purpose of this paper is to show an explicit proof for $d=1$.

We show the no-ghost theorem using the technique of Frenkel, Garland and
Zuckerman (FGZ) \cite{FGZ}. Their proof is different from the others. For
example, the standard BRST quantization assumes $d \geq 2$ in order to prove
the ``vanishing theorem," {\it i.e.}, the BRST cohomology is trivial except
at the zero ghost number. However, 
FGZ's proof of the vanishing theorem essentially does not require this as we will see later.
Establishing the no-ghost theorem for $d=1$ is then
straightforward by calculating the ``index" and ``signature" of the cohomology groups.

Thus, in a sense our result is obvious a priori, from Refs.~\cite{FGZ,LZ,FK}.
But it does not seem to be known well, so it is worth working out this point
explicitly in detail.

The organization of the present paper is as follows. First, in the next
section, we set up our notations and briefly review the BRST quantization of
string theory. In Section~\ref{sec:vanishing}, we will see a
standard proof
of the vanishing theorem and review why the standard no-ghost theorem cannot
be extended to $d<2$. Then in Section~\ref{sec:fgz}, we prove the
vanishing theorem due to FGZ, following Refs.~\cite{FGZ,LZ,FK}.
We use this result in Section~\ref{sec:no-ghost} to prove the no-ghost
theorem for $d=1$.

For the other attempts of the $d=1$ proof, see Section~\ref{sec:discussion}~(iv). Among them, Ref.~\cite{GM} considers the same kind of CFT as ours. However, the proof relies heavily on the proof of the flat $d=26$ string, and thus the proof is somewhat roundabout and is not transparent as ours. Moreover, our proof admits the extention to more general curved backgrounds [see Section~\ref{sec:discussion}~(v)].


\section{BRST Quantization}\label{sec:brst}

In this section, we briefly review the BRST quantization of string theory
\cite{KO,Big,BMP}. We make the following assumptions:
\begin{enumerate}
\item[(i).] Our world-sheet theory consists of $d$ free bosons
$ X^{\mu}\; (\mu = 0,\cdots ,d-1)$ with
signature $ (1, d-1) $ and a compact unitary CFT $ K $ of central charge $
c_{K} = 26-d $. Although we focus on the $d=1$ case below, the extension to $1 \leq d \leq 26$ is straightforward [Section~\ref{sec:discussion}~(i)].
\item[(ii).] 
We assume that $K$ is unitary and that its spectrum is discrete and bounded
below. Thus, all states in
$K$ lie in highest weight representations. The weight of highest weight
states should have $h^K>0$ from the Kac determinant; therefore, the
eigenvalue of $L_0^K$ is always non-negative.
\item[(iii).] The momentum of states is $k^{\mu} \neq 0$. [See
Section~\ref{sec:discussion}~(iii) for the exceptional case $k^{\mu}=0$.]
\end{enumerate}
Then, the total $L_{m}$ of the theory is given by $L_{m} = L_{m}^{X} +
L_{m}^{g} +
L_{m}^{K}$, where
\begin{eqnarray}
L_{m}^{X} &=& \frac{1}{2} \sum_{n=-\infty}^{\infty}
                : \alpha^{\mu}_{m-n} \alpha_{\mu,n} :, \\
L_{m}^{g} &=& \sum_{n=-\infty}^{\infty} (m-n) : b_{m+n} c_{-n} : - \delta_{m}.
\end{eqnarray}
Here,
\begin{equation}
[ \alpha^{\mu}_{m}, \alpha^{\nu}_{n} ] = m \delta_{m+n} \, \eta^{\mu\nu},
\qquad
\{ b_m, c_n \} = \delta_{m+n},
\end{equation}
and $\delta_m = \delta_{m,0}$. With the $d$-dimensional momentum $k^\mu$,
$ \alpha^{\mu}_{0} = \sqrt{2\alpha'} k^{\mu} $.

The ghost number operator $\hN$ counts the number of $c$ minus the
number of $b$ excitations:
\footnote{The ghost zero modes will not matter to our discussion. $\hN$ is related to the standard ghost number operator $ N^g $ as $N^g=\hN + c_0 b_0 - \frac{1}{2}$. Note that the operator $\hN$ is also normalized so that $\hN \ket{\downarrow}=0$.}
\begin{equation}
\hN	= \sum_{m=1}^{\infty} (c_{-m}b_{m} - b_{-m}c_{m})
        = \sum_{m=1}^{\infty} (N_{m}^{c} - N_{m}^{b}).
\end{equation}

We will call the total Hilbert space ${\cal H}_{total}$.
Recall that the physical state conditions are
\begin{equation}
Q \ket{\mbox{phys}} =0, \qquad b_{0} \ket{\mbox{phys}} =0.
\end{equation}
These conditions imply
\begin{equation}
0 = \{ Q, b_{0} \} \ket{\mbox{phys}} = L_{0} \ket{\mbox{phys}}.
\label{eq:L0cond}
\end{equation}
Thus, we define the following subspaces of ${\cal H}_{total}$:
\begin{subequations}
\begin{eqnarray}
{\cal H} &=& \{ \phi \in {\cal H}_{total}: b_{0} \phi = 0 \},\\
\hat{\cal H} &=& {\cal H}^{L_0} 
	= \{ \phi \in {\cal H}_{total}: b_{0} \phi = L_0 \phi =0 \}.
\end{eqnarray}
\end{subequations}
We will consider the cohomology
on $\hat{\cal H}$ since $Q$ takes $\hat{\cal H}$ into itself from
$\{Q,b_0\}=L_0$ and $[Q,L_0]=0$. The subspace ${\cal H}$ will be useful in
our proof of the vanishing theorem (Section~\ref{sec:fgz}).

The Hilbert space $\hat{\cal H}$ is classified according to mass eigenvalues. $\hat{\cal H}$ at a particular mass level will be often written as $\hat{\cal H}(k^2)$.
For a state $ \ket{\phi} \in \hat{\cal H}(k^2) $,
\begin{equation}
L_{0} \ket{\phi} = (\alpha' k^2+L_{0}^{int}) \ket{\phi} = 0,
\label{eq:on_shell}
\end{equation}
where $L_{0}^{int}$ counts the level number. 
One can further take an
eigenstate of $ \hN $ since $ [ L_{0}^{int}, \hN ] = 0 $.
$\hat{\cal H}$ is decomposed by the eigenvalues of $\hN$ as
\begin{equation}
\hat{\cal H} = \bigoplus_{n\in {\bf Z}}\hat{\cal H}^n.
\end{equation}

We define the raising operators as $\alpha^{\mu}_{-m}, b_{-m}, c_{-m},
x^{\mu}$ and $c_0$. The ground state in $\hat{\cal H}(k^2)$ is given by 
\begin{equation}
\ket{0;k} \otimes \ket{h^K} 
= e^{ik \cdot x} \ket{0;\downarrow} \otimes \ket{h^K},
\end{equation}
where $\ket{0;\downarrow}$ is the vacuum state annihilated by all lowering
operators and $\ket{h^K}$ is a highest weight state in $K$. Then, $ \hat{\cal H}(k^2) $ is written as
\begin{equation}
\hat{\cal H}(k^2) =
\left(
{\cal F}(\alpha^{\mu}_{-m}, b_{-m}, c_{-m};k) \otimes {\cal H}_{K}
\right)^{L_0}.
\end{equation}
Here, $*^{L_0}$ denotes the $L_0$-invariant subspace:
$ F^{L_0}= F \cap \mbox{Ker} L_0$.
A state in ${\cal H}_{K}$ is constructed by Verma modules of $K$.
The Fock space ${\cal F}(\alpha^{\mu}_{-m}, b_{-m}, c_{-m};k)$ is spanned by all states of the form
\begin{equation}
\ket{N;k} =
\prod_{\mu=0}^{d-1} \prod_{m=1}^{\infty}
\frac{(\alpha^{\mu}_{-m})^{N_{m}^{\mu}}}{\sqrt{m^{N_{m}^{\mu}} N_{m}^{\mu}!}}
\prod_{m=1}^{\infty} (b_{-m})^{N_{m}^{b}}
\prod_{m=1}^{\infty} (c_{-m})^{N_{m}^{c}} \ket{0;k},
\label{eq:general_state}
\end{equation}
where 
$ N_{m}^{\mu} = \frac{1}{m} \alpha^{\mu}_{-m} \alpha_{\mu,m} $ 
are the number operators for $ \alpha^{\mu}_{m} $.
In terms of the number operators,
\begin{equation}
L_{0}^{int}
= \sum_{m=1}^{\infty} m \Bigl( N_{m}^{b} + N_{m}^{c}
+ \sum_{\mu=0}^{d-1} N_{m}^{\mu} \Bigr) + L_{0}^{K} -1.
\label{eq:L0int}
\end{equation}

The BRST operator
\begin{equation}
Q = \sum_{m=-\infty}^{\infty} ( L_{-m}^{X} + L_{-m}^{K} ) c_{m}
        - \frac{1}{2} \sum_{m,n=-\infty}^{\infty}
        (m-n) : c_{-m} c_{-n} b_{m+n} : - c_{0}
\end{equation}
can be decomposed in terms of ghost zero modes as follows:
\begin{equation}
Q = \hat{Q} + c_{0} L_{0} + b_{0} M,
        \label{eq:hatq}
\end{equation}
where $ M= -2 \sum^{\infty}_{m=1} m c_{-m} c_{m} $ and  $ \hat{Q} $ is the
collection of the terms in $ Q $ without $ b_{0} $ or $ c_{0} $.
Using the above decomposition (\ref{eq:hatq}), for a state $\ket{\phi} \in
\hat{\cal H}$,
\begin{equation}
Q \ket{\phi} = \hat{Q} \ket{\phi}.
        \label{eq:relative}
\end{equation}
Therefore, the physical state condition reduces to
\begin{equation}
\hat{Q} \ket{\phi} = 0.
\end{equation}
Also, $\hat{Q}^2=0$ on $\hat{\cal H}$ from Eq.~\eq{relative}. Thus,
$\hat{Q}: \hat{\cal H}^n \rightarrow \hat{\cal H}^{n+1} $
 defines a BRST complex, which is called the {\it relative} BRST
complex. So, we can define $\hat{\cal H}_{c}, \hat{\cal H}_{e}
\subset
\hat{\cal H}$ by
\begin{equation}
\hat{Q} \hat{\cal H}_{c} = 0, \qquad
\hat{\cal H}_{e} = \hat{Q} \hat{\cal H},
\end{equation}
and define the relative BRST cohomology of $Q$ by
\begin{equation}
\hat{\cal H}_{obs}=\hat{\cal H}_{c}/\hat{\cal H}_{e}.
\end{equation}
In terms of the cohomology group,
$\hat{\cal H}_{obs}(k^2)=\oplus_{n \in {\bf Z}}\, H^n(\hat{\cal
H}(k^2), \hat{Q}(k))$.

Now, the inner product in ${\cal H}_{total}$ is given by
\begin{equation}
\bra{0,I;k} c_{0} \ket{0,I';k'}
        = (2 \pi)^d \delta^d(k-k') \delta_{I,I'},
\end{equation}
where $I$ labels the states of the compact CFT $K$. We take the basis $I$
to be orthonormal. Then, the inner product $\bra{}\ket{}$ in $\hat{\cal H}$
 is defined by $\langle | \rangle$ as follows:
\begin{equation}
\bra{0,I;k} c_{0} \ket{0,I';k'}
        = 2 \pi \delta (k^2-k'{}^2) \bra{0,I;k} \ket{0,I';k'}.
\end{equation}
The inner products of the other states follow from the algebra
with the hermiticity property, $b_{m}^\dagger = b_{-m},\, c_{m}^\dagger = c_{-m}$
and
$(\alpha^{\mu}_{m})^\dagger = \alpha^{\mu}_{-m}$.%
\footnote{We will write $\bra{\cdots}\ket{\cdots}$ as $\langle \cdots |
\cdots \rangle$ below.}

\section{The Vanishing Theorem and Standard Proofs}\label{sec:vanishing}

In order to prove the no-ghost theorem, it is useful to show the following
theorem:
\begin{theorem}[The Vanishing Theorem]
The $\hQ$-cohomology can be non-zero only at $ \hN =0 $, {\it i.e.}, 
$H^n(\hat{\cal H}, \hQ)=0$ for $n\neq 0$.
\label{thm:vanishing}
\end{theorem}
\noindent To prove this, we use the notion of {\it filtration}.
We first explain the method and then give an example of filtration
used in~\cite{KO,BMP}.
The filtration is part of the reason why $d \geq 2$ in standard proofs.

A filtration is a procedure to break up $\hQ$ according to a quantum number
$ N_{f} $ (filtration degree):
\begin{equation}
\hQ = Q_{0}+Q_{1}+\cdots+Q_{N},
\end{equation}
where
\begin{equation}
[N_{f}, Q_{m}] = m Q_{m}.
\end{equation}
We also require
\begin{equation}
[N_{f}, \hN] = [N_{f}, L_{0}] = 0.
\label{eq:cond}
\end{equation}
Then, $\hat{\cal H}$ breaks up according to the filtration degree $N_f (=q)$ as well as the ghost number $\hN (=n)$:
\begin{equation}
\hat{\cal H} = \bigoplus_{q,n\in {\bf Z}}\hat{\cal H}^{n;q} .
\end{equation}
If $\hat{\cal H}^q$ can be nonzero only for a finite range of degrees, the
filtration is called {\it bounded}.

The nilpotency of $ \hQ^2 $ implies
\begin{equation}
\sum_{\stackrel{\scriptstyle m, n}{m+n=l}}
Q_{m}Q_{n} = 0,  \qquad l=0, \ldots ,2N
\end{equation}
since they have different values of $ N_{f} $. In particular,
\begin{equation}
Q_{0}^2 = 0.
\end{equation}
The point is that we can first study the cohomology of
$ Q_{0} : \hat{\cal H}^{n;q}\rightarrow \hat{\cal H}^{n+1;q}$.
This is easier since $ Q_{0} $ is often simpler than $ \hQ$.
Knowing the cohomology of $ Q_{0} $ then tells us about the cohomology of $
\hQ $. In fact, one can show the following lemma:
\begin{lemma}
If the $ Q_{0} $-cohomology is trivial,  so is the $ \hQ
$-cohomology.
\label{lemma:exact_sequence}
\end{lemma}

\begin{proof}
\footnote{The following argument is due to Ref.~\cite{BMP}.}
Let $ \phi $ be a state of ghost number $ \hN =n$ and $\hQ$-invariant
($\phi\in \hat{\cal H}^n_c$).
Assuming that the filtration is bounded, we write
\begin{equation}
\phi = \phi_{k} + \phi_{k+1} + \cdots + \phi_{p},
\end{equation}
where $\phi_q\in \hat{\cal H}^{n;q}$.
Then,
\begin{equation}
\hQ \phi = (Q_{0} \phi_{k}) + (Q_{0} \phi_{k+1} + Q_{1} \phi_{k})
+  \cdots + (Q_N\phi_p).
\end{equation}
Each parenthesis vanishes separately since they carry different $ N_{f} $.
So, $ Q_{0} \phi_{k} =0 $.  The $ Q_{0} $-cohomology is trivial by
assumption, thus $ \phi_{k} = Q_{0} \chi_{k} $.
But then $ \phi
- \hQ \chi_{k} $, which is cohomologous to $ \phi $, has no $ N_{f} = k $
piece. By induction, we can eliminate all $ \phi_q $, so $ \phi = \hQ (
\chi_{k} + \ldots + \chi_{p} ) $; $ \phi $ is actually $\hQ$-exact.
\end{proof}

Moreover, one can show that the $Q_{0}$-cohomology is isomorphic to that of
$\hQ$ if the $Q_{0}$-cohomology is nontrivial for at most one filtration degree
\cite{Big,BMP}. We do not present the proof because our derivation does not
need this.
In the language of a spectral sequence \cite{BT}, the first term and the
limit term of the sequence are
\begin{equation}
E_1 \cong \bigoplus_q H(\hat{\cal H}^q, d_0), \qquad E_\infty \cong
H(\hat{\cal H}, \hQ).
\end{equation}
The above results state that the sequence collapses after the first term:
\begin{equation}
E_1 \cong E_\infty.
\end{equation}
Then, a standard proof proceeds to show that states in the nontrivial degree are in fact light-cone spectra, and thus there is no ghost in the $\hQ$-cohomology \cite{Big}.

Now, we have to find
an appropriate filtration and show that the $Q_0$-cohomology is trivial
if $\hN \neq 0$.
This completes the proof of the vanishing theorem.
The standard proof of the theorem  uses the
following filtration \cite{KO,BMP}:
\footnote{The $\hN$ piece is not really necessary.
We include this to make the filtration degree non-negative.}
\begin{equation}
N^{(KO)}_{f} = \sum_{\stackrel{\scriptstyle m=-\infty}{m \neq 0}}^{\infty}
        \frac{1}{m} \alpha^{-}_{-m} \alpha^{+}_{m} + \hN.
\label{eq:Nko}
\end{equation}
The degree $N^{(KO)}_{f}$ counts the number of $\alpha^+$ minus the number of
$\alpha^-$
excitations. So, this filtration assumes two flat directions.
The degree zero
part of $\hQ$ is
\begin{equation}
Q^{(KO)}_{0} = -\sqrt{2 \alpha'} k^{+}
        \sum_{\stackrel{\scriptstyle m=-\infty}{m \neq 0}}^{\infty}
        c_{m} \alpha^{-}_{-m}.
\end{equation}
The operator $Q^{(KO)}_{0}$ is nilpotent since $\alpha^{-}_{m}$ commute and
$c_{m}$ anticommute. Obviously, we cannot use $\alpha^{0}_{m}$ in place of
$\alpha^{-}_{m}$
since $\alpha^{0}_{m}$ do not commute. Thus,
we have to take a different approach for $d=1$.


\section{The Vanishing Theorem (FGZ)}\label{sec:fgz}

Since we want to show the no-ghost theorem for $d=1$, we cannot use $
N^{(KO)}_{f} $ as our filtration degree. Fortunately, there is a different
proof of the vanishing theorem \cite{FGZ,LZ,FK}, which uses a different
filtration.
Their filtration is unique in that $Q_0$ can actually be written as a sum of
two differentials, $d'$ and $d''$. This effectively reduces the problem to a
``$c=1$" CFT, which contains the timelike part and the $b$ ghost part. Then,
a K\"{u}nneth formula relates the theorem to the whole complex.
This is the reason why the proof does not require $d \geq 2$.
In addition, in this approach the
matter Virasoro generators themselves play a role similar to that of the
light-cone oscillators in Kato-Ogawa's approach. In this section, we 
prove the theorem using the technique of
Refs.~\cite{FGZ,LZ,FK}, but for more
mathematically rigorous discussion,
consult the original references.

\begin{proof}[Proof of the vanishing theorem for $d=1$]
FGZ's filtration is originally given for the $d=26$
flat spacetime as
\begin{equation}
N^{(FGZ)}_{f} = -L_{0}^{X(d=26)} + \sum_{m=1}^{\infty} m ( N_{m}^{c} -
N_{m}^{b} ).
\label{eq:fgz}
\end{equation}
The filtration itself does not require $d \geq 2$;
this filtration can be
naturally used for $d=1$, replacing $L_{0}^{X(d=26)}$ with
$L_{0}^{X}$. Then, the modified filtration assigns the following degrees to the operators:
\begin{subequations}
\begin{eqnarray}
&&\mbox{fdeg}(c_{m}) =  |m|, \qquad \mbox{fdeg}(b_{m}) = -|m|, \qquad \\
&&\mbox{fdeg}(L_{m}^{X}) = m, \qquad \mbox{fdeg}(L_{m}^{K}) = 0. 
\end{eqnarray}
\end{subequations}

The operator $N^{(FGZ)}_{f}$ satisfies conditions~\eq{cond} and the
degree of each term in $\hQ$ is non-negative. Because the eigenvalue of
$L_{0}^{int}$ is bounded below from Eq.~\eq{L0int}, the total number of
oscillators for a given mass level is bounded. Thus, the degree for the
states is bounded for each mass level. Note that the unitarity of the
compact CFT $K$ is essential for the filtration to be bounded.

The degree zero part of $\hQ$ is given by
\begin{subequations}
\label{eq:q0}
\begin{eqnarray}
Q^{(FGZ)}_{0} &=& d'+d'', \\
d' &=& \sum_{m>0} c_{m} L^{X}_{-m}
        + \sum_{m,n>0} \frac{1}{2} (m-n) b_{-m-n} c_{m} c_{n}, \\
d'' &=& - \sum_{m,n>0} \frac{1}{2} (m-n) c_{-m} c_{-n} b_{m+n}.
\end{eqnarray}
\end{subequations}
We break $\hat{\cal H}$ as follows:
\begin{subequations}
\begin{eqnarray}
\hat{\cal H} & = & \left({\cal F}(\alpha^{0}_{-m}, b_{-m}, c_{-m};k^0)
                                        \otimes {\cal H}_{K}\right)^{L_0} \\
        & = & \left({\cal F}(\alpha^{0}_{-m}, b_{-m};k^0) 
        \otimes {\cal F}(c_{-m})
\otimes {\cal H}_{K}\right)^{L_0}.
\end{eqnarray}
\end{subequations}
The Hilbert spaces $\hat{\cal H}$, ${\cal F}(\alpha^{0}_{-m}, b_{-m};k^0)$
and
${\cal F}(c_{-m})$ are decomposed according to the ghost number
$\hN=n$:
\begin{equation}
\hat{\cal H}^{n} =
\Biggl(
\Bigl(
\bigoplus_{\stackrel{\scriptstyle n=N^{c}-N^{b}}{N^{c}, N^{b} \geq 0}}
{\cal F}^{-N^{b}}(\alpha^{0}_{-m}, b_{-m};k^0) 
\otimes {\cal F}^{N^{c}}(c_{-m})
\Bigr) \otimes {\cal H}_{K}
\Biggr)^{L_0}.
\end{equation}
{}From Eqs.~\eq{q0}, the differentials act as follows:
\begin{subequations}
\begin{eqnarray}
&& Q^{(FGZ)}_{0}: {\cal H}^n \rightarrow {\cal H}^{n+1}, \\
&& d': {\cal F}^{n}(\alpha^{0}_{-m}, b_{-m};k^0) \rightarrow 
{\cal F}^{n+1}(\alpha^{0}_{-m}, b_{-m};k^0), \\
&& d'': {\cal F}^{n}(c_{-m}) \rightarrow {\cal F}^{n+1}(c_{-m}),
\end{eqnarray}
\end{subequations}
and $d'{}^2 = d''{}^2 = 0$. Thus, ${\cal F}^{n}(\alpha^{0}_{-m}, b_{-m};
k^0)$ and ${\cal F}^{n}(c_{-m})$ are complexes with differentials $d'$
and $d''$. Note that $Q^{(FGZ)}_{0}$ is the differential for ${\cal H}^n$ as
well as for $\hat{\cal H}^{n}$.

Then, the K\"{u}nneth
formula (Appendix~\ref{app:A}) relates the cohomology group of ${\cal H}$ to
those of ${\cal F}(\alpha^{0}_{-m}, b_{-m};k^0)$ and ${\cal F}(c_{-m})$:
\begin{equation}
H^{n}({\cal H}) =
\Biggl(
\bigoplus_{\stackrel{\scriptstyle n=N^{c}-N^{b}}{N^{c}, N^{b} \geq 0}}
                H^{-N^{b}}\left({\cal F}(\alpha^{0}_{-m}, b_{-m};k^0)\right)
                \otimes
                H^{N^{c}}\left({\cal F}(c_{-m}^{})\right)
\Biggr)
\otimes {\cal H}_K.
                \label{eq:kunneth}
\end{equation}
Later we will prove the following lemma:
\begin{lemma}
$ H^{-N^{b}}\left({\cal F}(\alpha^{0}_{-m}, b_{-m};k^0)\right) = 0 $ if
$N^{b} > 0$ and $(k^0)^2>0$.
\label{lemma4.1}
\end{lemma}
\noindent Then, Eq.~\eq{kunneth} reduces to
\begin{equation}
H^{n}({\cal H})^{L_0} =
                \left(
                \bigoplus_{n=N^{c} }
                H^{0}\left({\cal F}(\alpha^{0}_{-m}, b_{-m};k^0)\right)
                \otimes
                H^{N^{c}}\left({\cal F}(c_{-m}^{})\right)
                \otimes {\cal H}_K \right)^{L_0},
\end{equation}
which leads to $H^{n}({\cal H})^{L_0} = 0$ for $n < 0$ because $N^c\geq 0$.
The cohomology group $H^{n}({\cal H})^{L_0}$ is not exactly what we want.
However, Lian and Zuckerman have shown that
\begin{equation}
H^n(\hat{\cal H}) \cong H^n({\cal H})^{L_0}.
\end{equation}
See pages 325-326 of Ref.~\cite{LZ}. Thus,
\begin{equation}
H^{n}(\hat{\cal H}, Q^{(FGZ)}_{0}) = H^{n}(\hat{\cal H}, \hQ) = 0 
\cond{if $n<0$.}
\end{equation}
We will later prove the Poincar\'{e} duality theorem,
$ H^{n}(\hat{\cal H}, \hQ) = H^{-n}(\hat{\cal H}, \hQ) $
(Theorem~\ref{lemma:poincare}). Therefore,
\begin{equation}
H^{n}(\hat{\cal H}, \hQ) = 0 \cond{if $n \neq 0$.}
\end{equation}
This is the vanishing theorem for $d=1$.
\end{proof}

Now we will show Lemma~\ref{lemma4.1}. The proof is twofold: the first
is to
map the $c=1$ Fock space ${\cal F}(\alpha^{0}_{-m};k^0)$ to a Verma module, and the second is to show the lemma
using the Verma module. 

Let ${\cal V}(c, h)$ be a Verma module 
with highest weight $h$ and central charge~$c$. 
Then, we first show the isomorphism
\begin{equation}
{\cal F}(\alpha^{0}_{-m};k^0) \cong {\cal V}(1, h^{X}) 
\cond{if $(k^0)^2>0$.}
\label{eq:verma}
\end{equation}
Here, $h^{X}=-\alpha'(k^0)^2$. This is plausible from the defining formula of $L^{X}_{m}$,
\begin{equation}
L^{X}_{-m} = \sqrt{2\alpha'} k_0 \alpha^{0}_{-m} + \cdots,
\end{equation}
where $+ \cdots$ denotes terms with more than one oscillators. The actual 
proof is rather similar to an argument in \cite{thorn1,Brower:1971qr}.

\begin{proof}[Proof of Eq.~\eq{verma}]
The number of states of the Fock space ${\cal F}(\alpha^{0}_{-m};
k^0)$ and that of the Verma module ${\cal V}(1, h^{X})$ are the same for
a given level $N$. Thus, the Verma module%
\footnote{With a slight abuse of terminology, we use the word ``Verma module" even if $\ket{{h^{X},\{\lambda\}}}$ are not all independent.} 
furnishes a basis of the Fock
space if all the states in a highest weight representation,
\begin{equation}
\ket{{h^{X},\{\lambda\}}} =
L^{X}_{-\lambda_1}  L^{X}_{-\lambda_2} \ldots L^{X}_{-\lambda_M} \ket{h^{X}},
\end{equation}
are linearly independent, where $ 0 < \lambda_{1} \leq \lambda_{2} \leq
\cdots \leq \lambda_{M} $. This can be shown using the Kac determinant.

%
%

Consider the matrix of inner products for the states at level $N$:
\begin{equation}
{\cal M}_{\{\lambda\}, \{\lambda'\}}^{N}(c, h^{X}) =
\norm{h^{X},\{\lambda\}}{h^{X},\{\lambda'\}},
\qquad \sum_i \lambda_i = N.
\label{eq:matrix}
\end{equation}
The Kac determinant is then given by
\begin{equation}
\mbox{det} [{\cal M}^{N}(c, h^{X})] = K_N \prod_{1 \leq rs \leq N}
(h^{X}-h_{r,s})^{P(N-rs)},
\end{equation}
where $K_N$ is a positive constant and the multiplicity of the roots,
$P(N-rs)$, is the partition of $N-rs$. The zeros of the Kac determinant are
at
\begin{equation}
h_{r,s} = \frac{c-1}{24} + \frac{1}{4} (r \alpha_+ + s \alpha_-)^2,
\end{equation}
where
\begin{equation}
\alpha_{\pm} = \frac{1}{\sqrt{24}} (\sqrt{1-c} \pm \sqrt{25-c}).
\end{equation}
For $c=1$, $\alpha_{\pm} = \pm 1$ so that $ h_{r,s} = (r-s)^2/4 \geq 0 $.
Thus, the states $\ket{{h^{X},\{\lambda\}}}$ are linearly independent if $h^{X}<0$.
\end{proof}

Let us check what spectrum actually appears in $\hat{\cal H}$. Using assumption~(ii) of Section~\ref{sec:brst}, Eqs.~\eq{on_shell} and \eq{L0int},  we get $h^{X} \leq 1$ for a state in $\hat{\cal H}$. Also, $h^{X} \neq 0$ from assumption~(iii).%
\footnote{
The Verma module ${\cal V}(1, 0)$ fails to furnish the basis of ${\cal F}(\alpha^{0}_{-m};0)$ at the first level because $L^{X}_{-1} \ket{h^{X}=0} = 0$ for $d=1$.\label{ft:exceptional}}
Thus, we need to consider the Fock spaces ${\cal F}(\alpha^{0}_{-m};k^0)$ with $h^{X} \leq 1$ ($h^{X} \neq 0$). Those with $h^{X} <0$ are expressed by Verma modules from Eq.~\eq{verma}. On the other hand, those with $0 < h^{X} \leq 1$ are not. However, there is only the ground state in this region as the states in $\hat{\cal H}$. This state has $\hN=0$, so the state does not affect the vanishing theorem.

The isomorphism \eq{verma} is essential for proving the vanishing theorem.
In the language of FGZ, what we have shown is that ${\cal
F}(\alpha^{0}_{-m};k^0)$ is an ``${\cal L}_-$-free module," which is a prime
assumption of the vanishing theorem (Theorem~1.12 of \cite{FGZ}). The proof
of Lemma~\ref{lemma4.1} is now straightforward using 
Eq.~\eq{verma} and an argument given in \cite{FK}:

\begin{proof}[Proof of Lemma~\ref{lemma4.1}]
{}Using Eq.~\eq{verma},
a state $\ket{\phi} \in
{\cal F}(\alpha^{0}_{-m}, b_{-m};k^0)$ can be written as
\begin{equation}
\ket{\phi} = b_{-i_{1}} \ldots b_{-i_{L}}
                L^{X}_{-\lambda_{1}} \ldots L^{X}_{-\lambda_{M}}
\ket{h^{X}},
\label{eq:statebn}
\end{equation}
where $ 0 < \lambda_{1} \leq \lambda_{2} \leq \cdots \leq \lambda_{M} $ and
$ 0 < i_{1} < i_{2} < \cdots < i_{L} $.
Note that the states in ${\cal F}(\alpha^{0}_{-m}, b_{-m};k^0)$
all have nonpositive ghost number: $\hN \ket{\phi} = -L \ket{\phi}$.

We define a new filtration degree $N^{(FK)}_{f}$ as
\begin{equation}
N^{(FK)}_{f} \ket{\phi} = -(L+M) \ket{\phi},
\end{equation}
which corresponds to
\begin{equation}
\mbox{fdeg}(L^{X}_{-m}) = \mbox{fdeg}(b_{-m}) = -1
\cond{for $m>0$.}
\end{equation}
The algebra then determines $ \mbox{fdeg}(c_{m}) = 1 $ (for $m>0$)
from the assignment.
The operator $N^{(FK)}_{f}$ satisfies conditions~\eq{cond}
and the degree of each term in $d'$ is non-negative.
The degree zero part of $d'$ is given by
\begin{equation}
d'_{0} = \sum_{m>0} c_{m} L^{X}_{-m}.
\end{equation}
Since we want a bounded filtration, break up 
${\cal F}(\alpha^{0}_{-m}, b_{-m};k^0)$
according to $L_0$ eigenvalue $l_0$ :
\begin{equation}
{\cal F}(\alpha^{0}_{-m}, b_{-m};k^0) 
= \bigoplus_{l_0} {\cal F}(\alpha^{0}_{-m}, b_{-m};k^0)^{l_0},
\end{equation}
where
\begin{equation}
{\cal F}(\alpha^{0}_{-m}, b_{-m};k^0)^{l_0}
 = {\cal F}(\alpha^{0}_{-m}, b_{-m};k^0)
             \cap\mbox{Ker}(L_0-l_0).
\end{equation}
Note that ${\cal F}(\alpha^{0}_{-m}, b_{-m};k^0)^{l_0}$ is finite dimensional
since $ \ket{\phi} \in {\cal F}(\alpha^{0}_{-m}, b_{-m};k^0)^{l_0}$
satisfies
\begin{equation}
\sum^L_{k=1}i_k + \sum^M_{k=1}\lambda_k + h^{X} =l_0 .
\end{equation}
Thus, the above filtration
 is bounded for each ${\cal F}(\alpha^{0}_{-m}, b_{-m};k^0)^{l_0}$.

We first consider the $d'_{0}$-cohomology on
${\cal F}(\alpha^{0}_{-m}, b_{-m};k^0)^{l_0}$ for each $l_0$.
Define an operator $\Gamma$ such as
\begin{equation}
\Gamma \ket{\phi} =
\sum_{l=1}^{M} b_{-\lambda_{l}} \left( b_{-i_{1}} \ldots b_{-i_{L}} \right)
L^{X}_{-\lambda_{1}} \ldots \widehat{{L}^{X}_{-\lambda_{l}}} \ldots
L^{X}_{-\lambda_{M}} \ket{h^{X}},
\end{equation}
where $\widehat{{L}^{X}_{-\lambda_{l}}}$ means that the term is missing
(When $M=0$, $\Gamma \ket{\phi} \defi 0$). Then,
it is straightforward to show that
\begin{equation}
\{ d'_{0}, \Gamma \} \ket{\phi} = (L+M) \ket{\phi}.
\end{equation}
The operator $\Gamma$ is called a {\it homotopy operator} for $d'_{0}$. Its
significance is that the $d'_{0}$-cohomology is trivial except for $L+M =
0$. If $\ket{\phi}$ is closed, then
\begin{equation}
\ket{\phi}      = \frac{\{ d'_{0}, \Gamma \}}{L+M} \ket{\phi}
                = \frac{1}{L+M} d'_{0} \Gamma \ket{\phi}.
\end{equation}
Thus, a closed state $\ket{\phi}$ is actually an exact state if $L+M\neq 0$.
Therefore, the $d'_{0}$-cohomology is trivial if $\hN<0$ since $\hN=-L$.
And now, again using Lemma~\ref{lemma:exact_sequence},
the $d'$-cohomology $H^n({\cal F}(\alpha^{0}_{-m}, b_{-m};k^0)^{l_0})$
is trivial if $n<0$.

Because $[d',L_0]=0$, we can define
\begin{equation}
H^n({\cal F}(\alpha^{0}_{-m}, b_{-m};k^0))^{l_0} =
H^n({\cal F}(\alpha^{0}_{-m}, b_{-m};k^0))\cap \mbox{Ker}(L_0-l_0).
\end{equation}
Furthermore, the isomorphism
\begin{equation}
H^n({\cal F}(\alpha^{0}_{-m}, b_{-m};k^0))^{l_0} \cong
H^n({\cal F}(\alpha^{0}_{-m}, b_{-m};k^0)^{l_0})
\end{equation}
can be established. Consequently, 
$H^n({\cal F}(\alpha^{0}_{-m}, b_{-m};k^0))=0$
if $n <0$.
\end{proof}

\section{The No-Ghost Theorem}\label{sec:no-ghost}

Having shown the vanishing theorem, it is straightforward to show
the no-ghost theorem:
\begin{theorem}[The No-Ghost Theorem]
$\hat{\cal H}_{obs}$ is a positive definite space when $1 \leq d \leq 26$.
\end{theorem}
\noindent The calculation below is essentially the same as the one in
Refs.~\cite{FGZ,spiegelglas,FK}, but we repeat it here for completeness.

In order to prove the theorem, the notion of
{\it signature} is useful. 
For a vector space $V$ with an inner product,
we can choose a basis $e_{a}$ such that
\begin{equation}
\norm{e_{a}}{e_{b}} = \delta_{ab} C_{a},
\label{eq:basis}
\end{equation}
where $C_{a}\in \{0, \pm 1 \}$.
Then, the signature of $V$ is defined as
\begin{equation}
\sig{V} = \sum_{a} C_{a},
\end{equation}
which is independent of the choice of $e_{a}$.
Note that if $ \sig{V} = \dims{V} $, all the
$C_{a}$ are 1, so $V$ has positive definite norm.

So, the statement of the no-ghost theorem is equivalent to
\footnote{In this section, we also write
$V^{obs}_i = \hat{\cal H}_{obs}(k^2)$ and $V_i =\hat{\cal H}(k^2)$, where the subscript $i$ labels different mass levels.}
\begin{equation}
\sig{V^{obs}_{i}} = \dims{V^{obs}_{i}}.
        \label{eq:noghost1}
\end{equation}
This can be replaced as a more useful form
\begin{equation}
\sum_{i} e^{-\lambda \alpha' m_i^2} \sig{V^{obs}_{i}}
        = \sum_{i} e^{-\lambda \alpha' m_i^2} \dims{V^{obs}_{i}},
        \label{eq:noghost2}
\end{equation}
where 
$\lambda$ is a constant. 
Equation~\eq{noghost1} can be retrieved from Eq.~\eq{noghost2} by expanding in powers of $\lambda$. 
We write Eq.~\eq{noghost2} as
\begin{equation}
\mbox{tr}_{obs} \, e^{-\lambda L_{0}^{int}} C
        = \mbox{tr}_{obs} \, e^{-\lambda L_{0}^{int}},
\label{eq:noghost3}
\end{equation}
where $C$ is an operator which gives eigenvalues $C_{a}$.

Equation~\eq{noghost3} is not easy to calculate; however, the following
relation is straightforward to prove:
\begin{subequations}
\begin{equation}
\mbox{tr} \, e^{-\lambda L_{0}^{int}} C
        = \mbox{tr} \, e^{-\lambda L_{0}^{int}} (-)^{\hN}.
\label{eq:step3}
\end{equation}
Here, the trace is taken over $V_{i}$
and we take a basis which diagonalizes $(-)^{\hN}$.
Thus, we can prove Eq.~\eq{noghost3} by showing Eq.~\eq{step3} and
\begin{eqnarray}
\tr e^{-\lambda L_{0}^{int}} (-)^{\hN}
        &=& \mbox{tr}_{obs} \, e^{-\lambda L_{0}^{int}},
        \label{eq:step1} \\
\tr e^{-\lambda L_{0}^{int}} C
        &=& \mbox{tr}_{obs} \, e^{-\lambda L_{0}^{int}} C.
        \label{eq:step2}
\end{eqnarray}
\end{subequations}
Thus, the trace weighted by $(-)^{\hN}$ is an {\it index}.
 

\begin{proof}[Proof of Eq.~\eq{step1}]
At each mass level, states $\varphi_{m}$ in $V_{i}$ are classified
into two kinds of representations: BRST singlets $\phi_{\ta} \in V^{obs}_{i}$
and BRST
doublets $(\chi_{a}, \psi_{a})$, where $\chi_{a} = \hQ \psi_{a}$.
The ghost
number of $\chi_{a}$ is the ghost number of $\psi_{a}$ plus 1. Therefore,
$(-)^{\hN}$ causes these pairs of states to cancel in the index and only the singlets
contribute:
\begin{eqnarray}
\tr e^{-\lambda L_{0}^{int}} (-)^{\hN}
        &=& \mbox{tr}_{obs} \, e^{-\lambda L_{0}^{int}} (-)^{\hN} \\
        &=& \mbox{tr}_{obs} \, e^{-\lambda L_{0}^{int}}.
\end{eqnarray}
We have used the vanishing theorem on the last line.
\end{proof}

\begin{proof}[Proof of Eq.~\eq{step2}]
At a given mass level, the matrix of 
inner products among $\ket{\varphi_{m}}$ takes the form
\begin{equation}
\norm{\varphi_{m}}{\varphi_{n}}
=
\left(\begin{array}{c}
        \bra{\chi_{a}} \\
        \bra{\psi_{a}} \\
        \bra{\phi_{\ta}}
\end{array}\right)
\left( \ket{\chi_{b}}, \ket{\psi_{b}}, \ket{\phi_{\tb}} \right)
=
\left( \begin{array}{ccc}
        0 & M & 0 \\
        M^{\dagger} & A & B \\
        0 & B^{\dagger} & D
\end{array} \right).
\label{eq:matrixab}
\end{equation}
We have used $\hat{Q}^\dagger = \hat{Q}$,
$ \norm{\chi}{\chi} = \bra{\chi} \hQ \ket{\psi} = 0 $ and
$ \norm{\chi}{\phi} = \bra{\psi} \hQ \ket{\phi} = 0 $.
If $M$ were degenerate, there would be a state $\chi_{a}$ which is orthogonal to all states in $V_{i}$. Thus, the matrix $M$ should be nondegenerate. (Similarly, the matrix $D$ should be nondegenerate as well.) 
So, a change of basis
\begin{eqnarray}
\ket{\chi'_{a}} &=& \ket{\chi_{a}},
	\nonumber \\
\ket{\psi'_{a}} &=& \ket{\psi_{a}}
	- \frac{1}{2} (M^{-\dagger} A)_{ba} \ket{\chi_{b}},
	\nonumber \\
\ket{\phi'_{\ta}} &=& \ket{\phi_{\ta}}
	- (M^{-\dagger} B)_{b \ta} \ket{\chi_{b}},
\label{eq:transf1}
\end{eqnarray}
sets $A=B=0$. Finally, going to a basis,
\begin{eqnarray}
\ket{\chi''_{a}} &=&
	\frac{1}{\sqrt{2}} (\ket{\chi'_{a}}
	+ M^{-1}_{ba} \ket{\psi'_{b}}), \nonumber \\
\ket{\psi''_{a}} &=&
	\frac{1}{\sqrt{2}} (\ket{\chi'_{a}}
	- M^{-1}_{ba} \ket{\psi'_{b}}), \nonumber \\
\ket{\phi''_{\ta}} &=& \ket{\phi'_{\ta}},
\label{eq:transf2}
\end{eqnarray}
the inner product $\norm{\varphi''_{m}}{\varphi''_{n}}$ becomes
block-diagonal:
\begin{equation}
\norm{\varphi''_{m}}{\varphi''_{n}}
=
        \left( \begin{array}{cccccc}
                1 & 0   & 0 \\
                0 & -1 & 0 \\
                0 & 0   & D
        \end{array} \right).
\end{equation}
Therefore, BRST doublets again make no net contribution:
\begin{equation}
\tr e^{-\lambda L_{0}^{int}} C
        = \mbox{tr}_{obs} \, e^{-\lambda L_{0}^{int}} C.
\end{equation}
This proves Eq.~\eq{step2}.
\end{proof}

One can indeed check that $M$ and $D$ are nondegenerate. 
The inner product in $V_i$ is 
written as the product of inner products in  
${\cal F}(\alpha^{0}_{-m};k^0)$, ghost sector and ${\cal H}_K$.
The inner product in ${\cal F}(\alpha^{0}_{-m};k^0)$ is easily seen 
to be diagonal and nondegenerate.
For the ghost sector, 
the inner product 
becomes diagonal and nondegenerate as well by taking the
basis $ p_{m} = (b_{m}+c_{m})/\sqrt{2} $ and $ m_{m} =
(b_{m}-c_{m})/\sqrt{2} $, 
where
\begin{equation}
\{ p_m, p_n \} = \delta_{m+n}, \qquad
\{ p_m, m_n \} = 0, \qquad
\{ m_m, m_n \} = -\delta_{m+n}.
\label{eq:diagonal_basis}
\end{equation}
${\cal H}_K$ is assumed to have a positive-definite inner product.
Therefore, the matrix $\norm{\varphi_{m}}{\varphi_{n}}$ is nondegenerate.
Consequently, the matrices $M$ and $D$ are also nondegenerate. 

The inner product is
nonvanishing only between the states with opposite ghost numbers. 
Since $D$ is nondegenerate, BRST singlets of opposite ghost number must 
pair up.
We have therefore established the Poincar\'{e} duality
theorem as well:
\begin{theorem}[Poincar\'{e} Duality]
$ H^{\hN}(\hat{\cal H}, \hQ) = H^{-\hN}(\hat{\cal H}, \hQ) $.
\label{lemma:poincare}
\end{theorem}

\begin{proof}[Proof of Eq.~\eq{step3}]
We prove Eq.~\eq{step3} by explicitly calculating the both sides.

In order to calculate the left-hand side of Eq.~\eq{step3}, take an orthonormal basis of
definite $N^{p}_{m}$, $N^{m}_{m}$ [the basis 
\eq{diagonal_basis}], $N^{0}_{m}$ and an orthonormal basis of ${\cal
H}_{K} $. Then, $C=(-)^{N^{m}_{m} + N^{0}_{m}}$. Similarly, for the right-hand side, take an orthonormal basis of definite
$N^{b}_{m}$, $N^{c}_{m}$, $N^{0}_{m}$ and an orthonormal basis of ${\cal
H}_{K}$. 

%
%

{}From these relations, the left-hand side of Eq.~\eq{step3} becomes
\begin{eqnarray}
\lefteqn{\tr e^{-\lambda L_{0}^{int}} C} \nonumber \\
        &=& e^{\lambda} \prod_{m=1}^{\infty}
                \left( \sum_{N^{p}_{m}=0}^{1}
                        e^{-\lambda m N^{p}_{m}} \right)
                \left( \sum_{N^{m}_{m}=0}^{1}
                        e^{-\lambda m N^{m}_{m}} (-)^{N^{m}_{m}} \right)
                        \nonumber \\
        && \mbox{} \hspace{1.5in} \times \left( \sum_{N^{0}_{m}=0}^{\infty}
                        e^{-\lambda m N^{0}_{m}} (-)^{N^{0}_{m}} \right)
                \, \mbox{tr}_{{\cal H}_{K}} \, e^{-\lambda L_{0}^{K}}
                        \nonumber \\
        &=& e^{\lambda} \prod_{m}
                (1+e^{-\lambda m})(1-e^{-\lambda m})(1+e^{-\lambda m})^{-1}
                \, \mbox{tr}_{{\cal H}_{K}} \, e^{-\lambda L_{0}^{K}}
                        \nonumber \\
        &=& e^{\lambda} \prod_{m} (1-e^{-\lambda m})
                \, \mbox{tr}_{{\cal H}_{K}} \, e^{-\lambda L_{0}^{K}}.
\end{eqnarray}
The right-hand side becomes
\begin{eqnarray}
\lefteqn{\tr e^{-\lambda L_{0}^{int}} (-)^{\hN}} \nonumber \\
        &=& e^{\lambda} \prod_{m=1}^{\infty}
                \left( \sum_{N^{b}_{m}=0}^{1}
                        e^{-\lambda m N^{b}_{m}} (-)^{N^{b}_{m}} \right)
                \left( \sum_{N^{c}_{m}=0}^{1}
                        e^{-\lambda m N^{c}_{m}} (-)^{N^{c}_{m}} \right)
                        \nonumber \\
        && \mbox{} \hspace{1.5in} \times \left( \sum_{N^{0}_{m}=0}^{\infty}
                        e^{-\lambda m N^{0}_{m}} \right)
                \, \mbox{tr}_{{\cal H}_{K}} \, e^{-\lambda L_{0}^{K}}
                        \nonumber \\
        &=& e^{\lambda} \prod_{m}
                (1-e^{-\lambda m})(1-e^{-\lambda m})(1-e^{-\lambda m})^{-1}
                \, \mbox{tr}_{{\cal H}_{K}} \, e^{-\lambda L_{0}^{K}}
                        \nonumber \\
        &=& e^{\lambda} \prod_{m} (1-e^{-\lambda m})
                \, \mbox{tr}_{{\cal H}_{K}} \, e^{-\lambda L_{0}^{K}}.
\end{eqnarray}
This proves Eq.~\eq{step3}.
\end{proof}

\section{Discussion}\label{sec:discussion}

(i). The extension of the vanishing theorem to $d \geq 1$ is straightforward. 
Write $ \hat{\cal H} $ such that
\begin{equation}
\hat{\cal H}=
\left(
{\cal F}(\alpha^{0}_{-m}, b_{-m}, c_{-m};k^0) \otimes {\cal H}_{s}
\right)^{L_0},
\end{equation}
where
$ {\cal H}_{s} = {\cal F}(\alpha^{i}_{-m};k^i) \otimes {\cal H}_{K} $. The
superscript $i$ runs from 1 to $d \! - \! 1$. Similarly, break up $ L_{m} $. In particular,
\begin{subequations}
\begin{equation}
L_{0}=L^{(0)}_{0}+L^{g}_{0}+L^{(s)}_{0},
\end{equation}
where
\begin{eqnarray}
L^{(0)}_{0}+L^{g}_{0}
	&=& h^{(0)}
	+ \sum_{m=1}^{\infty} m ( N_{m}^{0}+N_{m}^{b}+N_{m}^{c}  ) - 1,
	\\
L^{(s)}_{0} &=& h^{(s)} 
	+ \sum_{i=1}^{d-1} \sum_{m=1}^{\infty} m N_{m}^{i} + L_{0}^{K},
	\\
h^{(0)} &=& -\alpha' (k^0)^2, \qquad
h^{(s)} = \sum_{i=1}^{d-1} \alpha' (k^i)^2.
\end{eqnarray}
\end{subequations}
Just like ${\cal H}_{K}$, the spectrum of ${\cal H}_{s}$ is bounded below and $L_{0}^{(s)} \geq 0$ for $(k^i)^2 \geq 0$.%
\footnote{In fact, FGZ have shown that an infinite sum of Verma modules with $h>0$ furnish a basis of the Fock space ${\cal F}(\alpha^{i}_{-m};k^i)$. Thus, not only ${\cal H}_{K}$, but the whole ${\cal H}_{s}$ must be written by Verma modules with $h>0$.}
Thus, our derivation applies to $d>1$ essentially with no modification; simply make the following replacements:
\begin{equation}
{\cal H}_{K} \rightarrow {\cal H}_{s},
L_{0}^{X} \rightarrow L_{0}^{(0)}, 
L_{0}^{K} \rightarrow L_{0}^{(s)},
h^{X} \rightarrow h^{(0)}.
\end{equation}
(However, use the only momentum independent piece of $L_{0}^{(s)}$ in calculating the index and the signature.)

(ii). The standard proofs of the no-ghost theorem 
do not only show the theorem, but also show
that the BRST cohomology is isomorphic to the
light-cone spectra. 
Since we do not have light-cone directions in general, we do not show this. In other words, we do not construct physical
states explicitly.

(iii). Our proof does not apply at the exceptional value
of momentum $k^{\mu}=0$ because the vanishing theorem fails (See footnote~\ref{ft:exceptional}). Even in the
flat $d=26$ case, the exceptional case needs a separate treatment
\cite{henneaux,FGZ,BMP}. 
For the flat case, the relative cohomology is nonzero at three ghost numbers and is represented by
\begin{equation}
\alpha^{\mu}_{-1} \ket{0;k^{\mu}=0},
\quad
b_{-1} \ket{0;k^{\mu}=0},
\quad \mbox{and} \quad
c_{-1} \ket{0;k^{\mu}=0}.
\end{equation}
Thus, there are negative norm states. However, the physical interpretation of these infrared states is unclear \cite{henneaux,Witten:1992yj}.

(iv). The original no-ghost theorem by Goddard and Thorn \cite{ocq2} can be applied to $d=1$ via a slight modification. The $d=1$ Hilbert space $ {\cal H} $ can be decomposed as
\begin{equation}
{\cal H}_{h^X,h^K} = 
\Bigl\{ Polynomial(\alpha^{0}_{-m}, L_{-m}^{K}) \ket{h^X, h^K} \Bigr\}.
\end{equation}
Goddard and Thorn's proof applies to any invariant subspace of the $d=26$ Hilbert space. The subspace ${\cal H}_{h^X,h^K}$ is invariant under the action of Virasoro generators. Moreover, any state of $K$ can be constructed by a free-field representation since $h^K>0$. Reference~\cite{GM} uses these facts to show the theorem for $d=1$. Incidentally, Thorn \cite{thorn1} also used OCQ and proved the no-ghost theorem for $1 \leq d \leq 25$. The proof does not assume the compact CFT, and there is no known way to give such a theory consistent interactions at loop levels \cite{Mandelstam:1974fq}.

(v). Finally, as is clear from our proof, the vanishing theorem itself does not  require even $d=1$, and the extention to more general backgrounds is possible. In particular, 
\begin{theorem}
$H^n(\hat{\cal H}, \hQ)=0$ for $n\neq 0$ if $ {\cal H} $ can be decomposed as
\begin{equation}
{\cal H}_{h,h'} = {\cal V}(c=1, h<0) \otimes {\cal V}(c'=25, h'>0) \otimes {\cal F}(b_{-m}, c_{-m}).
\end{equation}
\end{theorem}
\noindent Here, ${\cal V}(c=1, h<0)$ necessarily corresponds to a nonunitary CFT Hilbert space, and ${\cal V}(c'=25, h'>0)$ corresponds to a unitary CFT Hilbert space. Of course, we cannot prove the no-ghost theorem in our current technology. However, the no-ghost theorem implies the vanishing theorem; so, the proof of the vanishing theorem provides a consistency check or a circumstantial evidence of the no-ghost theorem for more general backgrounds.%
\footnote{work in progress}


\vspace{.1in}
\begin{center}
    {\Large {\bf Acknowledgments} }
\end{center}
\vspace{.1in}

We would like to thank N. Ishibashi, M. Kato, M. Sakaguchi, Y. Satoh and C.
Thorn for useful discussions. We would especially like to thank J.
Polchinski for various discussions and suggestions; the possibility of
showing the theorem for $d=1$ was pointed out by him to the authors.
The work of M.A. was supported in part by JSPS Research Fellowship for 
Young Scientists. The work of M.N. was supported in part by the
Grant-in-Aid for Scientific Research (11740161) from the Ministry of
Education, Science and Culture, Japan.

\appendix

\section{The K\"{u}nneth Formula}\label{app:A}

To simplify the notation, denote the complexes appeared in Section~\ref{sec:fgz} as follows:
\begin{subequations}
\begin{eqnarray}
({\cal F}(\alpha^{0}_{-m}, b_{-m}, c_{-m};k^0), Q^{(FGZ)}_{0}) 
& \rightarrow & ({\cal F},Q), \\
({\cal F}(\alpha^{0}_{-m},b_{-m};k^0), d') & \rightarrow & ({\cal F}_1,d'), \\
({\cal F}(c_{-m}), d'') & \rightarrow & ({\cal F}_2,d'').
\end{eqnarray}
\end{subequations}

Let $\{ \omega_i^{-b} \}$ and $\{ \eta_j^{n+b} \}$ be bases of $H^{-b}({\cal F}_1)$ and $H^{n+b}({\cal F}_2)$ respectively. Then, $\phi^n = \omega^{-b}_i
\eta^{n+b}_j$ is a closed state in ${\cal F}\,(={\cal F}_1\otimes{\cal F}_2)$.
 We show that this is not an
exact state. If it were exact, it would be written as
\begin{equation}
\phi^n = \omega_i^{-b} \eta_j^{n+b} = Q (\alpha^{-b-1} \beta^{n+b} +
\gamma^{-b} \delta^{n+b-1})
\end{equation}
for some $\alpha^{-b-1}$, $\beta^{n+b}$, $\gamma^{-b}$, and
$\delta^{n+b-1}$. Executing the differential, we get
\begin{eqnarray}
\omega_i^{-b} \eta_j^{n+b} &=&
(d' \alpha^{-b-1}) \beta^{n+b} + \alpha^{-b-1} (-)^{b+1} (d'' \beta^{n+b})
\nonumber \\
&& + (d' \gamma^{-b}) \delta^{n+b-1} + \gamma^{-b} (-)^b (d'' \delta^{n+b-1}).
\end{eqnarray}
Comparing the left-hand side with the right-hand side, we get $\alpha^{-b-1}
= \delta^{n+b-1} = 0$; thus, $\phi^n=0$ contradicting our assumption. Thus,
$\phi^n$ is an element of $H^{n}({\cal F})$. Conversely, any element of
$H^{n}({\cal F})$ can be decomposed into a sum of a product of the elements
of $H^{-b}({\cal F}_1)$ and $H^{n+b}({\cal F}_2)$. Thus, we obtain
\begin{equation}
H^{n}({\cal F}) =
\bigoplus_{\stackrel{\scriptstyle n=c-b}{c, b \geq 0}}
H^{-b}({\cal F}_1) \otimes H^{c}({\cal F}_2 ).
\end{equation}
Here, the restriction of the values $b$ and $c$ comes from the fact
${\cal F}^n_1={\cal F}^{-n}_2=0$ for $n>0$.

This is the K\"{u}nneth formula (for the ``torsion-free" case \cite{BT}.) Our discussion here is close to the one of Ref.~\cite{nakahara} for the de Rham cohomology.

\section{Some Useful Commutators}\label{app:C}

In this appendix, we collect some useful commutators:

\begin{align}
&[L_{m}, L_{n}] = (m-n) L_{m+n} + \frac{c}{12} (m^3-m) \delta_{m+n},
&& \nonumber \\
&[L_{m}, \alpha^{\nu}_{n}] = -n \alpha^{\nu}_{m+n}, 
&&[L_{m}, b_{n}] = (m-n) b_{m+n}, \nonumber \\
&[L_{m}, c_{n}] = (-2m-n) c_{m+n}, 
&& \nonumber \\
&[Q, L_{m}] = 0, 
&&[Q, \alpha^{\nu}_{m}] = - \sum_{n=-\infty}^{\infty} m c_{n} \alpha^{\nu}_{m-n}, \nonumber \\
&\{ Q, b_{m} \} = L_{m}, 
&&\{ Q, c_{m} \} = - \sum_{n=-\infty}^{\infty} n c_{-n} c_{m+n}, \nonumber \\
&[N^{g}, b_{m}] = -b_{m}, 
&&[N^{g}, c_{m}] = c_{m}, \nonumber \\
&[N^{g}, L_{m}] = 0,
 &&[N^{g}, Q] = Q. \nonumber
\end{align}

\end{document}